\documentclass[aps,prl,twocolumn,amsmath,amssymb,nofootinbib,superscriptaddress,floatfix,reprint,longbibliography]{revtex4-1}
\usepackage[dvips]{graphicx}
\usepackage{latexsym}
\usepackage{amsmath}
\usepackage{amsfonts}
\usepackage{amssymb}
\usepackage{color}
\usepackage{txfonts}
\usepackage{float}
\usepackage{url}
\usepackage[colorlinks=true, urlcolor=blue, linkcolor=blue, citecolor=blue]{hyperref}
\usepackage{ulem}
\usepackage{tikz-feynman}
\usepackage{tikz,tikz-feynhand} 
\usepackage{graphicx}
\usepackage{extpfeil}
\usepackage{subfigure}
\usepackage{physics}
\begin{document}
	\newcommand{\fig}[2]{\includegraphics[width=#1]{#2}}
	\newcommand{\la}{{\langle}}
	\newcommand{\ra}{{\rangle}}
	\newcommand{\dg}{{\dagger}}
	\newcommand{\upa}{{\uparrow}}
	\newcommand{\dna}{{\downarrow}}
	\newcommand{\ab}{{\alpha\beta}}
	\newcommand{\ias}{{i\alpha\sigma}}
	\newcommand{\ibs}{{i\beta\sigma}}
	\newcommand{\hH}{\hat{H}}
	\newcommand{\hn}{\hat{n}}
	\newcommand{\hc}{{\hat{\chi}}}
	\newcommand{\hU}{{\hat{U}}}
	\newcommand{\hV}{{\hat{V}}}
	\newcommand{\br}{{\bf r}}
	\newcommand{\bk}{{{\bf k}}}
	\newcommand{\bq}{{{\bf q}}}
	\def\gsim{~\rlap{$>$}{\lower 1.0ex\hbox{$\sim$}}}
	\setlength{\unitlength}{1mm}
	\newcommand{{\vhf}}{$\chi^\text{v}_f$}
	\newcommand{{\vhd}}{$\chi^\text{v}_d$}
	\newcommand{{\vpd}}{$\Delta^\text{v}_d$}
	\newcommand{{\ved}}{$\epsilon^\text{v}_d$}
	\newcommand{{\vved}}{$\varepsilon^\text{v}_d$}
	\newcommand{\pprl}{Phys. Rev. Lett. \ }
	\newcommand{\pprb}{Phys. Rev. {B}}

\title {Correlated BCS wavefunction approach to unconventional superconductors}
%%% \title {Unconventional properties in Correlated BCS wavefunction}
\author{Pengfei Li}
\affiliation{Beijing National Laboratory for Condensed Matter Physics and Institute of Physics,
	Chinese Academy of Sciences, Beijing 100190, China}
\affiliation{School of Physical Sciences, University of Chinese Academy of Sciences, Beijing 100190, China}

\author{Kun Jiang}
\email{jiangkun@iphy.ac.cn}
\affiliation{Beijing National Laboratory for Condensed Matter Physics and Institute of Physics,
	Chinese Academy of Sciences, Beijing 100190, China}
\affiliation{School of Physical Sciences, University of Chinese Academy of Sciences, Beijing 100190, China}

\author{Jiangping Hu}
\email{jphu@iphy.ac.cn}
\affiliation{Beijing National Laboratory for Condensed Matter Physics and Institute of Physics,
	Chinese Academy of Sciences, Beijing 100190, China}
\affiliation{Kavli Institute of Theoretical Sciences, University of Chinese Academy of Sciences,
	Beijing, 100190, China}
 \affiliation{New Cornerstone Science Laboratory, 
	Beijing, 100190, China}

\date{\today}

\begin{abstract}
We propose a modified BCS wavefunction as the ground state of a correlated superconductor with the correlation specified between $k$ and $-k$ electrons in the reciprocal space. Owing to this correlation, low-energy excitations are not conventional BCS Bogoliubov quasiparticles. They display at least four poles in the Green's function and their particle-hole weights in the tunneling and photon-emission spectrum become asymmetric. The superfluid density or superfluid stiffness also deviates from the BCS predictions with finite paramagnetic terms inside the total diamagnetic response. Moreover, a $d$-wave correlated pairing state becomes robust against weak disorders. Hence, this state can explain many mysterious features observed in unconventional superconductors like cuprates.
\end{abstract}
%\pacs{}
\maketitle

%\textit{Introduction}
The discovery of high-temperature superconductivity in cuprates \cite{bednorz}, whose transition temperatures greatly exceed other conventional superconductors (SCs), is one of the major achievements in condensed matter physics in the late twenty century \cite{doping_mott,keimer_review,armitage_RevModPhys.82.2421,Norman_2003_review}. Since its discovery in 1986, the research in cuprate superconductors greatly boosts the search for unconventional superconductors, like iron-based superconductors \cite{iron1,iron2,iron_review}, and nickel-based superconductors \cite{lidanfeng,meng_wang}.
Besides their high transition temperatures, the unconventional properties also enrich our view on quantum many-body physics. 
Owing to the development of modern techniques, including angle-resolved photoemission spectroscopy (ARPES) \cite{Shen_RevModPhys.75.473,Shen_RevModPhys.93.025006}, scanning tunneling microscope (STM)\cite{STM_RevModPhys.79.353,pan_stm,pan_stm2}, neutron scattering \cite{Neutron_PhysRevLett.58.2802,Neutron2,neutron_PhysRevLett.75.316}, we now have a new framework of unconventional superconductivity from different perspectives. For example, in cuprates, we have found the pseudogap phase with Fermi arc \cite{pesudogap_PhysRevLett.76.4841,pesudogap_optical1,pesudogap_optical2,ding_pseudogap}, the strange metal phase with linear-$T$ resistivity \cite{strange1,varma_RevModPhys.92.031001}, which are also intertwined with charge density waves, pair density waves, stripe phase, and other possible symmetry breaking phases \cite{fradkin}. Theoretically, how to arrive at this complicated phase diagram is a great challenge in condensed matter physics. Owing to the last thirty years' effort, many interesting theories have been proposed including resonating valence bonds (RVBs) \cite{RVB,Anderson_2004,anderson_history}, spin fluctuations \cite{scalapino_RevModPhys.84.1383}, SO(5) \cite{so5,so5_RevModPhys.76.909}, loop current \cite{varma_PhysRevB.73.155113,varma_RevModPhys.92.031001}, phase fluctuation \cite{phase_fluctuation}, intertwined orders \cite{fradkin} etc.
However, the strong correlation nature of unconventional superconductors makes the physics of high-temperature superconductivity remain highly debated. The underlying mechanism is becoming one of  the ``Holy Grails'' in physics research.

For most unconventional superconductors, it is widely believed that electron-electron correlation plays an essential role in their novel properties. On the other hand, this correlation always leads the quantum systems to another highly entangled regime lacking theoretical understanding and treatments.
A natural question arises here, can the BCS wavefunction faithfully describe the ground state wavefunction for unconventional SC in such a strongly correlated system \cite{bcs_theory,schrieffer}? 
Interestingly, even though the normal states of unconventional SCs like the cuprates are quite complicated without coherent description owing to strong correlation, the superconductivity is quite "normal`` with coherent quasiparticles from ARPES, STM, and other experimental measurements \cite{Shen_RevModPhys.75.473,Shen_RevModPhys.93.025006,Norman_2003_review,ding_PhysRevLett.87.227001,Norman_PhysRevLett.79.3506,pan_stm,STM_RevModPhys.79.353}. 
This feature suggests that there still has a coherent description for unconventional superconductors.
In this work, we use a correlated BCS (CBCS) wavefunction for unconventional superconductivity, which provides a unique way of understanding correlated quantum many-body superconductors.

\begin{figure}
	\begin{center}
		\fig{3.4in}{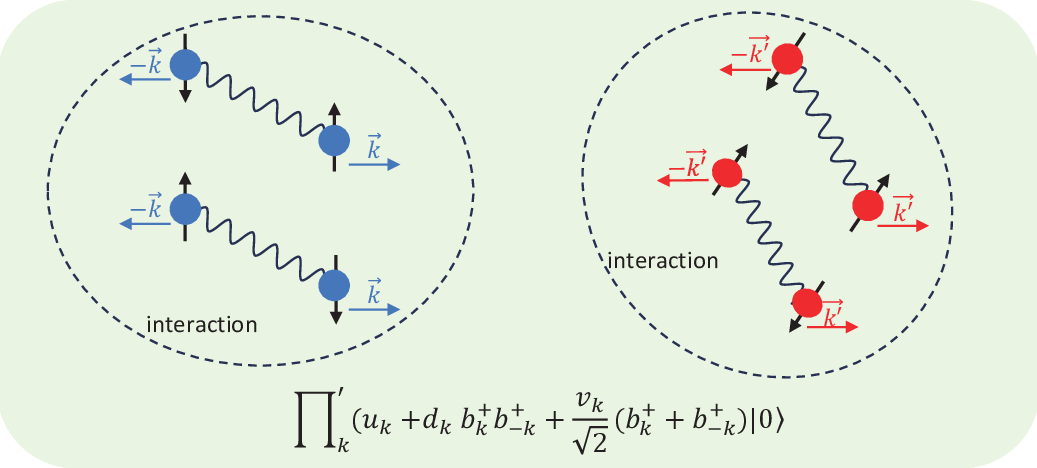}
		\caption{Schematic diagram of the correlated BCS wavefunction $|GS \rangle$. The cooper pairs created by $b_k^\dagger$ and $b_{-k}^\dagger$ are coupled by their correlation.
			\label{fig1}}
	\end{center}
	%	\vskip-0.5cm
\end{figure}

To describe any quantum systems, there are two essential ingredients: the ground state wavefunction, and their low-energy excitations. To keep the translation symmetry and the coherent features, we propose that the coherent part of ground state wavefunction $|GS \rangle$ for SCs takes the following form
\begin{eqnarray}
	|GS \rangle={\displaystyle \prod_k}'(u_k+d_kb_k^{\dagger}b_{-k}^{\dagger}+\frac{v_k}{\sqrt{2}}(b_k^{\dagger}+b_{-k}^{\dagger}))|0\rangle
\end{eqnarray}
where $b_k^{\dagger}=c_{k\uparrow}^\dagger c_{-k\downarrow}^\dagger$ is the pairing operator, as illustrated in Fig.\ref{fig1}. $u_k$, $d_k$, and $v_k$ are phenomenological parameters. ${\displaystyle \prod_k}'$ is restricted to half-Brillouin zone, since $k$ and $-k$ belong to the same Hilbert space.
Secondly, the excitations above $|GS \rangle$ have more poles than BCS excitations. Owing to the energy particle-hole symmetry, the retarded Green's function at least contains four poles with a following form
\begin{eqnarray}
	G^{R}(k,\omega)&=&\frac{(a_{k+})^2}{\omega-E_{+0k}+i\eta}+\frac{(a_{k-})^2}{\omega-E_{-0k}+i\eta} \\ \nonumber
	&&+\frac{(a_{k+}')^2}{\omega+E_{+0k}+i\eta}+\frac{(a_{k-}')^2}{\omega+E_{-0k}+i\eta}+G_{inc}
\end{eqnarray}
where $\pm E_{\pm0 k}$ are pole energies and $\eta \rightarrow 0^+$ for retarded Green's function. $a_{k\pm}$ and $a_{k\pm}'$ are also phenomenalogical parameters.  $G_{inc}$ is the background incoherent part.
The anomalous Green's functions $F(k,\omega)$ also take a four poles structure as discussed later. We find these two basic assumptions can capture many important features of unconventional SCs including particle-hole asymmetry, the missing superfluid density, impurity effects etc.

\subsection{Hilbert space, Wavefunction and Green's function}
Since the SC state remains coherent as discussed above, we believe that the $|GS \rangle$ remains a product state.
One of the simplest extensions of BCS wavefunction  ${\displaystyle \prod_k}(u_k+v_k b_k^{\dagger})|0\rangle$ starts from the local Hilbert space at each $k$ point to keep the translation symmetry. 
Owing to the fact that the pairing operator $b_k^{\dagger}$ connects the $k$ and $-k$, there are 16 possible states within each $k$ Hilbert space as listed in Table.\ref{table}. The dropping of NMR Knight shift after $T_c$ \cite{nmr_review,NMR_PhysRevB.56.877} eliminates the spin-triplet possiblity like $c_{k\uparrow}^\dagger c_{-k\uparrow}^\dagger  |0\rangle$ states.
$|GS \rangle$ should be also close to BCS, which rules out the odd electron states like $c_{k\uparrow}^\dagger|0\rangle$ or $c_{k\uparrow}^\dagger c_{k\downarrow}^\dagger c_{-k\uparrow}^\dagger  |0\rangle$. It is also highly unlike for the single $k$ fully occupied $c_{k\uparrow}^\dagger c_{k\downarrow}^\dagger  |0\rangle$ states. Therefore, the only possible combination is the mixing between $|0\rangle$, $\frac{1}{\sqrt{2}}(b_k^{\dagger}+b_{-k}^{\dagger}))|0\rangle$ and $b_k^{\dagger}b_{-k}^{\dagger}|0\rangle$. A natural generation ground wavefunction for correlated SC becomes
\begin{eqnarray}
	|GS \rangle={\displaystyle \prod_k}'(u_k+d_kb_k^{\dagger}b_{-k}^{\dagger}+\frac{v_k}{\sqrt{2}}(b_k^{\dagger}+b_{-k}^{\dagger}))|0\rangle
\end{eqnarray}
This CBCS wavefunction reduces to the BCS wavefunction if the ($|k\uparrow,-k\downarrow \rangle$, $|-k\uparrow,k\downarrow \rangle$) subspaces are decoupled with product state $(u_k+v_k b_k^{\dagger})(u_{-k}+v_{-k} b_{-k}^{\dagger})|0\rangle$.

 \begin{table}[htb]
 	\begin{tabular}{|c|c|c|}
 		\hline
 		subspace & state basis & basis number \\ \hline\hline
 		spin singlet & $|0\rangle$, $\frac{1}{\sqrt{2}}(b_k^{\dagger}+b_{-k}^{\dagger}))|0\rangle$ , $b_k^{\dagger}b_{-k}^{\dagger}|0\rangle$  & 3 \\ \hline
 	   spin triplet & $\frac{1}{\sqrt{2}}(b_k^{\dagger}-b_{-k}^{\dagger}))|0\rangle$, $c_{k\uparrow}^\dagger c_{-k\uparrow}^\dagger  |0\rangle$, $c_{k\downarrow}^\dagger c_{-k\downarrow}^\dagger  |0\rangle$ & 3  \\ \hline
 	  odd electron	 & \begin{tabular} {cc}
 	  	  $c_{k\uparrow}^\dagger|0\rangle$, &  $b_{-k}^\dagger c_{k\uparrow}^\dagger |0\rangle$ $\rightarrow$ $|O_{k+}^1\rangle$,$|O_{k-}^1\rangle$  \\ 
 	  	  $c_{k\downarrow}^\dagger |0\rangle$, & $b_{k}^\dagger  c_{k\downarrow}^\dagger |0\rangle$ $\rightarrow$ $|O_{k+}^2\rangle$,$|O_{k-}^2\rangle$ \\
 	  	  $c_{-k\uparrow}^\dagger |0\rangle$, &  $b_{k}^\dagger  c_{-k\uparrow}^\dagger |0\rangle$ $\rightarrow$ $|O_{k+}^3\rangle$,$|O_{k-}^3\rangle$ \\
 	  	  $c_{-k\downarrow}^\dagger |0\rangle$, &  $b_{-k}^\dagger c_{-k\downarrow}^\dagger |0\rangle$ $\rightarrow$ $|O_{k+}^4\rangle$,$|O_{k-}^4\rangle$ \\
 	  \end{tabular}
 	  & 8 \\ \hline
 	  k-single 	& $c_{k\uparrow}^\dagger c_{k\downarrow}^\dagger  |0\rangle$, $c_{-k\uparrow}^\dagger c_{-k\downarrow}^\dagger  |0\rangle$  & 2 \\  \hline
 	\end{tabular}
 	\caption{The local Hilbert space sectors at each k point with their basis number. The spin-singlet subspace is the central subspace we focus on. The odd electron subspace contains the most important low-energy excitations. We also list the eigenstates under $U_k$ in odd electron subspace after the $\rightarrow$ symbols. }
        \label{table}
 	\vskip-0.3cm
 \end{table}
 
%\textit{Hamiltonian}
Equipped with this wavefunction, we need to construct an effective Hamiltonian, whose ground state is this eigenstate. 
The only possible interaction without mixing $|GS\rangle$ wavefunction to other sectors are
$\hat{n}_{k\uparrow}\hat{n}_{k\downarrow}$, $\hat{n}_{k}\hat{n}_{-k}$, $\hat{S}_k\hat{S}_{-k}$.

\begin{eqnarray}
	H&=&\sum_{k}\epsilon_k c_{k\sigma}^{\dagger} c_{k\sigma}+U_{k} \hat{n}_{k\uparrow}\hat{n}_{k\downarrow}+U'_k\hat{n}_{k}\hat{n}_{-k}+J_k\hat{S}_k\hat{S}_{-k} \nonumber \\
	&&-(\Delta_k b_k^{\dagger}+\Delta_k b_k)+H_{res}
 \label{hk}
\end{eqnarray}
$\epsilon_k$ is the normal state dispersion. To include pairings, we introduce a phenomenological coupling $\Delta_k$. The origin of this pairing term is beyond the scope of this paper.
All these interactions $U_{k}$, $U_{k}'$, and $J_k$ should be considered as the residue interactions in the forward scattering channels, which are not renormalized towards zero under renormalization group flow. $H_{res}$ part is the other residue interaction or term related to incoherent parts of self-energy, which are not essential for our discussion. We will first ignore $H_{res}$ and discuss their impacts later. 
To further simplify our discussion and identify the central physics, we will focus on the $U_{k}\neq 0$, $U'_k=J_k=0$. Interestingly, this $U_{k}$ interaction model is just the Hatsugai–Kohmoto (HK) model \cite{HK_model,phillips_np}. Recently, this HK model has been widely studied to show non-Fermi liquid behaviors and unconventional superconductivity \cite{phillips_np,phillips_prb,Phillips_np2,fuchun_hk,Yangkun_PhysRevB.103.024529,yin_PhysRevB.106.155119}.
J. Zhao et al. also found that the HK model is a stable quartic fixed point in the presence of perturbing local interactions \cite{Phillips_RG}.

Hence, the spin singlet block Hamiltonian becomes
\begin{eqnarray}
	&&h_{G}(k)=\left(\begin{array}{ccc} 
		0 &- \sqrt{2}\Delta_k & 0 \\
		- \sqrt{2}\Delta_k & 2\epsilon_k & -\sqrt{2}\Delta_k\\
		0 & - \sqrt{2}\Delta_k & 4\epsilon_k+2U_k
	\end{array}\right) 
\end{eqnarray}
The ground state energy and wavefunction can be obtained numerically.

%\textit{Excitation and Green's function}
After constructing the $|GS\rangle$ wavefunction and $H$, we need to find the excitation states and Green's functions. The most important excited states connecting with $|GS\rangle$ are $odd$-electron states. For example, acting single-electron operators to the $|GS\rangle$, we have
\begin{eqnarray}
    c_{k\uparrow}|GS\rangle&=&(\frac{v_{k}}{\sqrt{2}}c_{-k\downarrow}^{\dagger}+d_{k}c_{-k\downarrow}^{\dagger} b_{-k}^{\dagger}) |GS'\rangle \label{excitation1}\\
    c_{k\uparrow}^\dagger|GS\rangle&=&(u_{k}c_{k\uparrow}^\dagger+\frac{v_{k}}{\sqrt{2}}c_{k\uparrow}^\dagger  b_{-k}^{\dagger})|GS'\rangle \label{excitation2}
\end{eqnarray}
if k is in the half BZ. $|GS'\rangle$ is the other part wavefunction without k. One new feature here is the electron operator applying to the $|GS\rangle$ leads to the mixing between another electron operator and a three-electron operator, which is different from the BCS case. Other excitation states can be obtained in the same way as discussed in the supplemental material (SM).  Then, the $c_{-k\downarrow}^{\dagger}$ and $c_{-k\downarrow}^{\dagger} b_{-k}^{\dagger}$ state are mixed through interactions and pairing. In our simplified discussion, only pairing terms mix state bases within each Hilbert subspace owing to the fact that the excited state bases are eigenstates of density-density interactions. For example, $c_{k\uparrow}^\dagger|0\rangle$ and $b_{-k}^\dagger c_{k\uparrow}^\dagger |0\rangle$ are coupled by the block Hamiltonian
\begin{eqnarray}
h_{odd}(k)=\left(\begin{array}{cc} 
		 \epsilon_k & -\Delta_k\\
		-\Delta_k & 3\epsilon_k+U_k
	\end{array}\right) 
 \label{hodd}
\end{eqnarray}
resulting in two excited states $|O_{k+}^1\rangle$, $|O_{k-}^1\rangle$ with energy $E_{\pm}=2\epsilon_k+\frac{U_k}{2}\pm E_{k}^{o}$ and $E_{k}^{o}=\sqrt{(\epsilon_k+\frac{U_k}{2})^2+\Delta_k^2}$. The other odd excited states are listed in Table \ref{table}.
In the following discussion, we will only focus on the ground state properties and we can write the electron operators via Hubbard operators connecting $|GS\rangle$ to $odd$-electron states.
For example, if we consider the $k$ inside our half-BZ, the electron operators can be written as
\begin{eqnarray}
c_{k\uparrow}^\dagger&=&
   a_{k+}^1 |O_{k+}^1\rangle \langle GS|+a_{k-}^1 |O_{k-}^1\rangle \langle GS|  \nonumber \\
   &+&a_{k+}^4 |GS\rangle \langle O_{k+}^4|+a_{k-}^4|G\rangle \langle O_{k-}^4|
   \label{c_dagger}
\end{eqnarray}
where $a_{k\pm}^{\alpha}$ is the $c_{k\uparrow}$ weight in each eigenstate.
Other operators are listed in the SM.
%if $k<0$
%\begin{eqnarray}
%c_{k\uparrow}^\dagger&=&
%   p_{3+k} |O_{k+}^3\rangle \langle G|+p_{3-k} |O_{k-}^3\rangle \langle G|  \nonumber \\
%   &+&p_{2+k} |G\rangle \langle O_{k+}^2|+p_{2-k}|G\rangle \langle O_{k-}^2|
%\end{eqnarray}

Then, Green's functions can be calculated using the Lehmann representation \cite{bruus2004many,coleman2015introduction}. We start from 
the single particle retarded Green's function $G^R_{\sigma}(k,t)=-i\theta(t)\langle \{c_{k \sigma}(t),c_{k\sigma}^\dagger(0)\}\rangle_{G}$. After the Fourier transformation, we obtain
\begin{eqnarray}
G_{\uparrow}(k,\omega)&=& \frac{ (a_{k+}^1)^2}{w-E_{+0}+i\eta}+\frac{(a_{k-}^1)^2}{w-E_{-0}+i\eta} \nonumber \\
	&+&\frac{(a_{k+}^4)^2}{w+E_{+0}+i\eta}+\frac{(a_{k-}^4)^2}{w+E_{-0}+i\eta}
 \label{green}
\end{eqnarray}
where $E_{\pm0}=E_{\pm}(k)-E_{G}$. From Eq.\ref{green}, we can easily find that there are four poles at energies $\pm E_{\pm0}$ as we claimed above.

We want to add a note here. The key physics discussed here is beyond the effective Hamiltonian Eq.\ref{hk}, and only depends on the $|GS\rangle$ and the low energy excitations or Green's functions. Since we don't have enough information to extract the weights and other coefficients from experiments, we will use this effective Hamiltonian Eq.\ref{hk} to illustrate the physics of correlated pairing.

\begin{figure}
	\begin{center}
		\fig{3.4in}{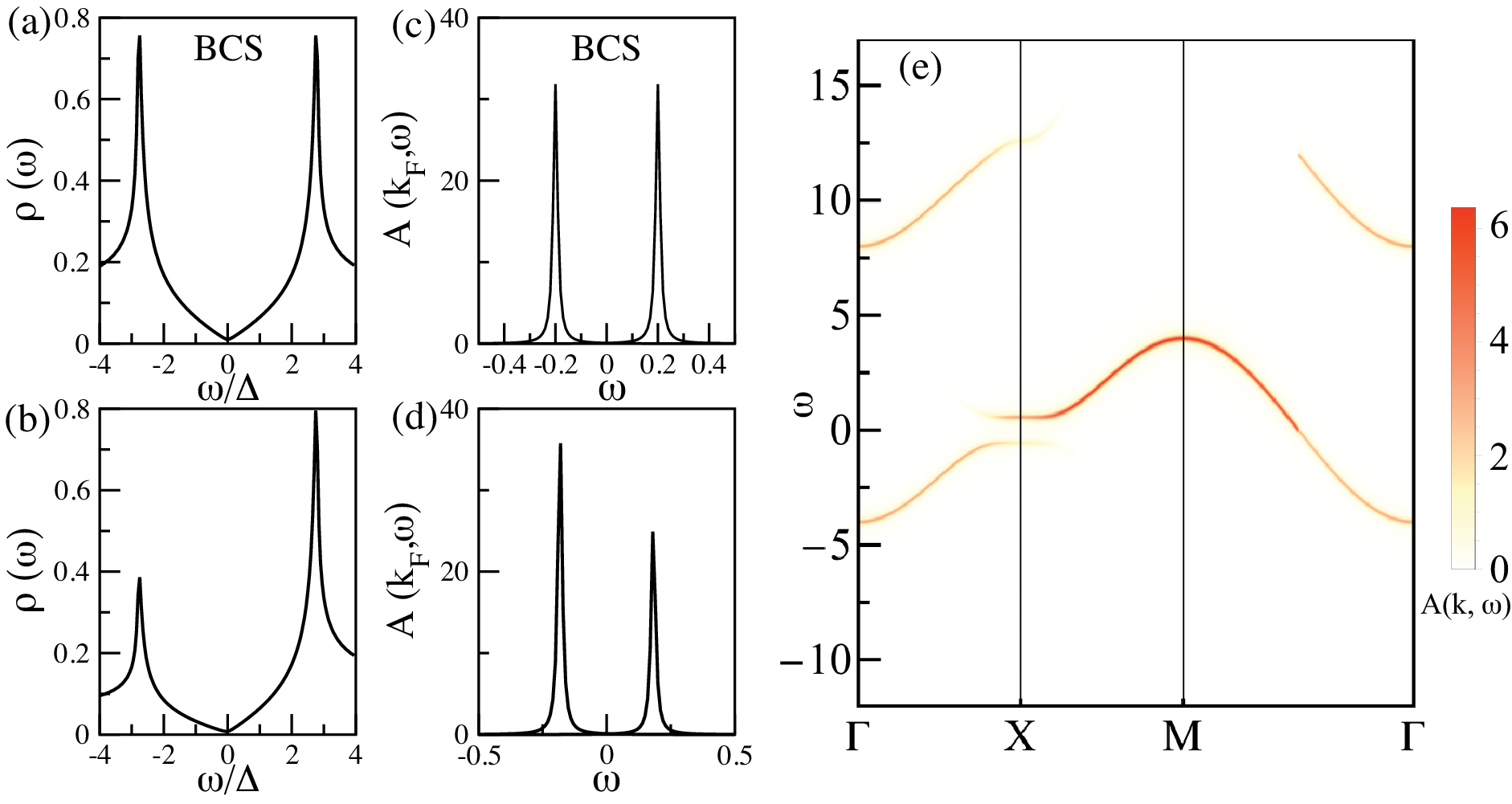}
		\caption{(a) The DOS $\rho(\omega)$ of the BCS state with $d$-wave pairing symmetry. (b) The DOS $\rho(\omega)$ from the correlated pairing model with $U_k=12$, $\Delta_0=0.2$. To remove the normal state DOS influence, we tune the chemical potential $\mu$ to the Van-Hove filling.
  (e) The corresponding $A(k,\omega)$ is also plotted. The $A(k_F,\omega)$ spectrum is asymmetric at the Fermi level.
  (c) The $A(k_F,\omega)$ spectrum of 12-sites 1D nearest-neighbor fermionic chain with extended $s$-wave pairing function $\Delta_0\cos k_x$ and $\Delta_0=0.2$. The chemical potential is tuned to $1.0$ for $k_F=2\pi/3$. (d) The $A(k_F,\omega)$ spectrum of the same 1D chain with Hubbard $U\hat{n}_{i\uparrow}\hat{n}_{i\downarrow}$ at each site $i$, where $U=0.2$. To obtain this spectrum, we implement the exact diagonalization code using the QuSpin package \cite{quspin1,quspin2}. More details can be found in SM.
			\label{fig2}}
	\end{center}
	%	\vskip-0.5cm
\end{figure}

\subsection{Particle-hole Asymmetry}
From Green's function in Eq.\ref{green}, we can infer that the excitation energies remain particle-hole symmetric while the spectrum weight may be different. To see this, we can calculate the density of states (DOS) $\rho(\omega)$ and the spectrum function $A(k,w)=-\frac{1}{\pi}Im G^R(k,\omega)$. To calculate these properties, the normal state dispersion in the square lattice is used with $\epsilon_k=-2t(\cos k_x+\cos k_y)-\mu$. We have set $t=1$ as the energy unit and used $\Delta(k)=\Delta_0(\cos k_x-\cos k_y)$ to characterize the d-wave pairing symmetry. The $\rho(\omega)$ obtained from Eq.\ref{green} with $U_k=12$, $\Delta_0=0.2$, $\mu=0$ is plotted in Fig.\ref{fig2} (b).
Interestingly, the two coherent peaks in $\rho(\omega)$ become weight-asymmetric. This asymmetric feature deviates from the BCS symmetric DOS plotted in Fig.\ref{fig2} (a).
In unconventional SC, it is always thought that the asymmetric structure in the tunneling spectrum comes from the bare band structure. This asymmetric feature is more evident when the bare band is around the van Hove filling. However, Anderson pointed out the band structure influence is canceled out due to Harrison's theorem \cite{harrison}. He concluded that this asymmetric feature inherits from the cuprates' correlated Mott nature \cite{anderson_history,anderson2016words}.

This asymmetric feature found here actually originates from the asymmetric spectrum weight in the particle-hole channel, which is evident in the $A(k_F,\omega)$ at the Fermi momentum $k_F$.
BCS theory tells us that its coherence factors have the same amplitudes at the Fermi level $E_F$. Hence, $A(k_F,\omega)$ is exactly symmetric from the BCS. However, this property is no longer valid under correlation. The calculated $A(k,\omega)$ from Eq.\ref{green} reveals the particle-hole weight is not symmetric, as shown in Fig.\ref{fig2}(e). This particle-hole asymmetric feature indicates that adding a particle and adding a hole to the ground state have different weights. This phenomenon has also been discussed in the Gutzwiller-RVB theory \cite{anderson_tunneling} and Gutzwiller approximation \cite{Randeria_PhysRevLett.95.137001}, which is also proposed to be one key feature of cuprates SCs by Anderson \cite{anderson_history}.

Actually, this asymmetric is more general beyond our analytical approach. Using exact diagonalization, we calculate the $A(k_F,\omega)$ in a 1D chain with extended $s$-wave pairing up to 12 sites. The particle-hole symmetry $A(k_F,\omega)$ is broken when we turn on the Hubbard interaction $U\hat{n}_{i\uparrow}\hat{n}_{i\downarrow}$, as plotted in Fig.\ref{fig2} (c-d).
All these findings tell us that for a correlated pairing system, the particle-hole weights are not needed to be equal although the corresponding energies are related. This asymmetry property roots in the asymmetry of ground state wavefunction factors $u_k$, $d_k$, and $v_k$. As in Eq. \ref{excitation1} and Eq. \ref{excitation2} respectively, particle excitations and hole excitations take similar forms. If the $c^{\dagger}b_{-k}$ terms have much higher energies, we should also expect $c_{-k\downarrow}^{\dagger}|GS'\rangle$ and $c_{k\uparrow}^{\dagger}|GS'\rangle$ lead to the same spectrum weight for a time-reversal invariant system. Hence, the particle-hole asymmetry is related to the asymmetry in their factors from $|GS\rangle$.

\subsection{Superfluid density}
After obtaining the asymmetric spectrum, we want to address the correlation-influenced superfluid density issue in cuprates SC.
The superfluid density is one key feature of superconductivity, which determines its diamagnetic Meissner effect ability \cite{schrieffer}. The diamagnetic responses in cuprates SC play an important role in the understanding of high-temperature superconductors
\cite{xiang2022d,Uemura2,ivan_nature,invan_review,Armitage_PhysRevLett.122.027003,Hirschfeld_PhysRevResearch.2.013228,Qianghua_PhysRevLett.128.137001,zixiang}.
%It is pointed out that.
%from the linear-doping dependence in under-doped samples to recently found $T_c$ scaling in overdoped samples

\begin{figure}
	\begin{center}
		\fig{3.4in}{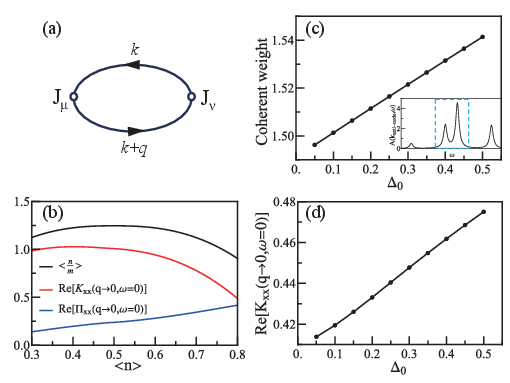}
		\caption{(a) The Feynman diagram of the paramagnetic current-current correlation function $\Pi_{\mu\nu}$. (b) The superfluid density (red line), paramagnetic response (blue line) and diamagnetic response (black line) in our cases with $d$-wave pairing symmetry at different filling conditions $\langle \hat{n} \rangle$. (c) The coherent weight (integrated spectral weight of the lower bands) at anti-node point with different $d$-wave pairing gap size $\Delta_0$. The inset is the schematic diagram of the spectrum weight distribution at anti-node point, in which the integration of the part circled by a dashed box is the cohenrent weight.  (d) The superfluid density with different $\Delta_0$ as (c). The parameters in the calculation are $\mu=2$, $U=12$. 
			\label{fig3}}
	\end{center}
	%	\vskip-0.5cm
\end{figure}

Theoretically, the superfluid density $n_s$ can be obtained from the linear response theory between current $J$ and gauge field $A$ \cite{xiang2022d,bruus2004many,Scalapino_PhysRevLett.68.2830}.
\begin{eqnarray}
    J_{\mu}=-\sum_{\nu}K_{\mu\nu}A_{\nu}=\left(-e^{2}\right)\sum_{\nu}\left[ \Pi_{\mu \nu}+\langle\frac{n}{m}\rangle \delta_{\mu \nu}\right] A_{\nu}.
    \label{ns}
\end{eqnarray}
where $K_{\mu\nu}$ is the current response function and $\mu,\nu=x,y,z$ for the spatial direction. $\langle\frac{n}{m}\rangle$ is the diamagnetic contribution. The $\Pi_{\mu \nu}$ is the paramagnetic current-current correlation function, which is defined as the Kubo formula 
\begin{eqnarray}
    \Pi_{\mu \nu}(q,\omega)=-\frac{i}{V\hbar}\int_{0}^{\infty}dt e^{i\omega t} \langle [J_\mu(q,t),J_\nu(-q,0)]\rangle.
\end{eqnarray}
and can be represented by the Feynman diagram in Fig.\ref{fig3}(a).
From the BCS theory \cite{schrieffer}, the $\Pi_{\mu \nu}$ contribution is zero owing to the SC gap. Hence, the $K_{\mu\mu}$ is determined by the diamagnetic term $\langle\frac{n}{m}\rangle$ leading to the London equation. From here, we can find that all the electrons contribute to the superfluid stiffness in the BCS theory. However, this behaviour is strongly challenged by recently found $n_s-T_c$ scaling in overdoped cuprates \cite{ivan_nature,armitage_RevModPhys.82.2421,invan_review} and previous found linear-doping dependence in underdoped cuprates \cite{Uemura_PhysRevLett.62.2317,Uemura2}.

For the tight-binding model on a lattice, the current operator \cite{Scalapino_PhysRevLett.68.2830} is defined as
\begin{eqnarray}
    J_\mu(q)=\sum_k \frac{\partial \epsilon_{k+\frac{q}{2}}}{\partial k_\mu} c_{k+q}^\dagger c_k
    \label{current}
\end{eqnarray}
The diamagnetic term $\langle\frac{n}{m}\rangle$ should be changed to \cite{xiang2022d,Scalapino_PhysRevLett.68.2830}
\begin{eqnarray}
    \langle\frac{n}{m}\rangle=\frac{1}{V}\sum_k\langle\frac{\partial^2 \epsilon_k}{\partial k_\mu\partial k_\nu}\rangle
\end{eqnarray}
One important thing here is the superfluid property is defined in the limit $K_{\mu\nu}(q\rightarrow0,\omega=0)$ \cite{Scalapino_PhysRevLett.68.2830,schrieffer,parks2018superconductivity}, which describes the response related to a static long-wavelength magnetic field. 

Normally, the $\langle\frac{n}{m}\rangle$ is easily calculated while the current-current correlation $\Pi_{\mu \nu}$ is subtle. In a clean $d$-wave superconductor,
the $\Pi_{\mu \nu}$ tends to zero at zero temperature as predicted from BCS theory. Hence, $K_{\mu\nu}(q\rightarrow0,\omega=0)$ comes from the diamagnetic part. The particle-hole excitation generated by $J_\mu(q\rightarrow0)$ is strictly zero in a clean conventional SC. 
This zero $\Pi_{\mu \nu}$  is similar to the vanishing matrix element from the optical selection rule \cite{optical_nagaosa}.
On the other hand, a dirty $d$-wave superconductor holds a finite DOS at the Fermi level from disorder scattering. These finite low-energy excitations
lead to finite contributions in the paramagnetic $\Pi_{\mu \nu}$. These disorder effects and possible granular pairing regions have been proposed to be reasons for the unconventional superfluid stiffness in the overdoped cuprates \cite{Hirschfeld_PhysRevResearch.2.013228,Qianghua_PhysRevLett.128.137001,zixiang}.

Interestingly, the paramagnetic $\Pi_{\mu \nu}(q\rightarrow0,\omega=0)$ from our model is also finite although the DOS $\rho(\omega=0)$ remains zero.
As shown in Fig.\ref{fig3}(b), we calculate the $K_{xx}(q\rightarrow0,\omega=0)$ as a function of doping $\langle n\rangle$ with fixed pairing strength $\Delta_0$. Although the $K_{xx}(q\rightarrow0,\omega=0)$ follows the similar trend in the paramagnetic term $\langle\frac{n}{m}\rangle$, the $\Pi_{xx}$ remains finite at zero temperature.
In our cases, the matrix element between the particle-hole excited states like $|O_{k-}^1O_{k+}^4\rangle$ and $|GS\rangle$ is nonzero while the overlapping between $|O_{k-}^1O_{k+}^1\rangle$ and $|GS\rangle$ remains zero. Therefore, $\Pi_{\mu \nu}(q\rightarrow0,\omega=0)$ is not vanishing, which leads to a paramagnetic contribution to $K_{\mu\nu}$. Although we still have nonvanishing superfluid density, the superfluid density still deviates from its BCS value at $T=0$. This behavior is consistent with our previous work in charge-4$e$ SC \cite{charge4e}. The reduction of superfluid density is both due to the existence of multiple poles in the Green's function caused by interactions, leading to non-zero transition matrix elements between different excited states, except that the interaction arises from quartet pairing in that case. Actually, in multi-band BCS superconductors, there is also similar result of the reduction of superfluid density \cite{Carbotte}.

More than that, we also find that superfluid stiffness $K_{xx}$ also depends on the pairing field $\Delta_0$, as plotted in Fig.\ref{fig3}(c). For the conventional BCS theory, superfluid stiffness only depends on the diamagnetic contribution, which is mainly a bare band value. There is no pairing field dependence in BCS. The pairing dependence of $K_{xx}$ mainly comes from the paramagnetic $\Pi_{xx}$. $\Pi_{xx}$ decreases as $\Delta_0$ increases inducing a increasing in $K_{xx}$.
Another important property that depends on $\Delta_0$ is the coherence peak weight around the Fermi level. As plotted in Fig.\ref{fig3}(d),
we found that this coherence weight is also linearly on $\Delta_0$. The combination of these two facts may be the reason for the link between the spectral weight and superfluid density found in ARPES before \cite{feng_science}.

\subsection{Impurity scattering}
%Impurity scattering is another important effect in superconductors, which also provides hallmarks for the $d$-wave pairing symmetries in cuprates SCs %\cite{balatsky2006impurity,xiang2022d,Hirschfeld_RevModPhys.81.45}. 
Based on BCS theory, the $d$-wave superconductor is quite sensitive to impurity scatterings because of the existence of nodes \cite{balatsky2006impurity}, which deviates from the Anderson theorem in conventional $s$-wave superconductors. Impurities serve as pair breakers producing low-energy excitations and finite weights around the pairing nodes.
On the other hand, the $d$-wave pairing in cuprates is quite robust in such a dirty complex oxide system \cite{keimer_review}. Commonly, there are two reasons: (1) most impurities locate off the CuO$_2$ planes in the charge reservoir layers; (2) the coherence length is very short. 
It has been proposed that the strong correlation is the reason for this insensitivity from a projected wavefunction study \cite{garg2008}.

\begin{figure}
	\begin{center}
		\fig{3.4in}{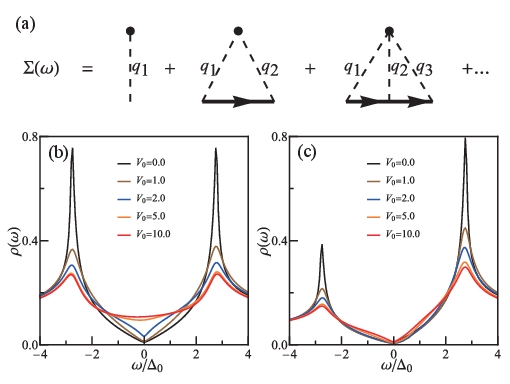}
		\caption{(a) The Feynman diagram of the self-energy with impurity scattering in T-martix approximation. (b) The DOS of BCS ground state with $d$-wave pairing symmetry corrected by impurity scattering. (c) The DOS in our model with $d$-wave pairing symmetry corrected by impurity scattering. The parameters in calculation are $\mu=0$, $U=2t$, $\Delta_0=0.2t$.
			\label{fig4}}
	\end{center}
	%	\vskip-0.5cm
\end{figure}

To explore this, we carry out a multiple-impurity scattering study from Hamiltonian in Eq.\ref{hk}. A standard $\delta$-function impurity scattering potential $U(r)=V_0 \delta(r)$ is insert.  When the density of impurities is dilute, we often neglect the interference between different impurities \cite{xiang2022d}. Then, the self-energy $\Sigma(\omega)$ from impurity scattering is obtained from the Feynman diagram plotted in Fig.\ref{fig4}(a). A self-consistent equation between $\Sigma(\omega)$ and Green's function is numerically calculated to find disorder-averaged DOS.

In order to compare with the solution from standard BCS theory, we plot the multi-impurity scattering results for a BCS $d$-wave pairing in Fig.\ref{fig4}(b). As the impurity strength $V_0$ increases, the coherent peaks keep dropping with a finite DOS $\rho(0)$ becoming obvious after $V=1.0$. This feature is consistent with the previous impurity pair breaking feature \cite{balatsky2006impurity,xiang2022d}. The small V region is close to the Born limit where $\rho(0)$ is finite but close to zero. 

On the contrary, the impurity scattering results in our case show a different behavior when we take the correlations into account. As plotted in Fig.\ref{fig4}(c), the coherent peaks keep dropping while $\rho(0)$ retains zero when $V_0$ increases. This feature directly proves that the disorder effect is suppressed in the strongly correlated superconductor. The nodal quasi-particle is protected by the interaction against impurity scatterings. Using Eq.\ref{hk}, we can also control the correlation range in momentum space. If we release correlation along the nodal direction, the impurity scattering will recover the standard $d$-wave results in Fig.\ref{fig4}(b). 

\subsection{Discussion and Summary}
Finally, we want to discuss how to go beyond our approach. We still want to emphasize that although we used a $k$ decoupled model in Eq.\ref{hk}, the physics above only depends on two things: the \textit{ground state wavefunction} and its \textit{low-energy excitation} behaviors. 
More precisely,
\begin{enumerate}
    \item The ground state wavefunction form relies on the product state assumption, which roots from the coherence peaks observed in ARPES, STM, and other techniques below $T_c$. For example, one consensus of cuprate superconductors is their $d$-wave pairing symmetry \cite{xiang2022d}. The phase-sensitive measurements using corner Josephson junctions and tricrystal Josephson rings provide concrete demonstrations for the $d$-wave pairing symmetry \cite{Harlingen_RevModPhys.67.515, Tsuei_RevModPhys.72.969}. This $d$-wave pairing feature has been widely tested from thermodynamic measurements, phase-sensitive junctions, superfluid density etc.\cite{xiang2022d}. Hence, the coherent description close to BCS wavefunction is justified. 
    \item The low-energy excitations stem from the assumption that the low-energy states are no longer single-particle states owing to correlation. In BCS theory or mean-field based on BCS theory, the low-energy states are also single-particle states related to $c_{k\sigma}^\dagger$ single-particle operators through Bogoliubov transformation. Then $c_{k\sigma}^\dagger$ is mapped to two particle-hole symmetric states. However, this is invalid here which implies $c_{k\sigma}^\dagger$ is mapped to at least four particle-hole symmetric states as in Eq.\ref{c_dagger} (more detailed discussion is listed in SM). Hence, Green's functions at least contain four poles as in Eq. \ref{green}.
\end{enumerate}
These two assumptions are direct consequences of electron correlation.

%is $\gamma_k=c_k\dagger+c_k \dagger b_{-k}$ 
%$\gamma=c_\dagger+c b_k$, which is contradict to the $\gamma=c_\dagger+c_{-k}$. $c_\dagger= $. 

Using two assumptions, we have demonstrated that the particle-hole weights in $A(k,\omega)$ and DOS are no longer symmetric while their energies remain symmetric. The particle-hole asymmetry is from the ground-state wavefunction asymmetry. Secondly, we find that the superfluid density at $T=0$ no longer only depends on the diamagnetic response function. The paramagnetic response function is also nonzero at $T=0$, which comes from the multipole Green's functions and
non-single-particle low-energy excitations. Then, the impurity scattering in our correlated wavefunctions has also been explored from self-consistent T matrix methods. Contrary to the well-established pair-breaking effects, the electron-electron correlation suppresses the non-magnetic impurity scattering effects. This provides another view on why $d$-wave paring is robust in cuprates.

In our toy model, we focus on the momentum decoupled interaction in Eq.\ref{hk} to achieve a solvable limit. Going beyond this solvable limit, we need to treat the residue Hamiltonian $H_{res}$. This $H_{res}$ can be treated as the spirit of Landau's Fermi liquid theory with quasiparticle interactions $f(k,k')$ and Landau parameters $F_{l}$ in different $l$ angular-momentum channels. Then, the $H_{res}$ plays a similar role of $f(k,k')$, which renormalizes the physical properties, induces self-energy and $G_{inc}$ background to Green's functions. These details don't change the conclusions in the current work and will be left for further exploration.
Hence, although the interactions in Eq.\ref{hk} imply infinite range interactions from Fourier transformation, the remaining $H_{res}$ is important for cutting the interactions range. But due to the limited phase space, $H_{res}$ plays a similar role of $f(k,k')$ in Fermi liquids.

In summary, we propose a wavefunction approach to unconventional superconductors from the ground state wavefunction and its unconventional excitations. Our new approach provides a unique way for correlated pairing, unveiling many puzzles in correlated superconductors. We hope our findings could stimulate the investigation of correlated superconducting ground states and their glamorous phenomena.

\subsection{Acknowledgement}
We thank Prof. Fu-Chun Zhang for the helpful and insightful discussions.
This work is supported by the Ministry of Science and Technology  (Grant No. 2022YFA1403901), the National Natural Science Foundation of China (Grant No. NSFC-11888101, No. NSFC-12174428), the Strategic Priority Research Program of the Chinese Academy of Sciences (Grant No. XDB28000000, XDB33000000), the New Cornerstone Investigator Program, and the Chinese Academy of Sciences through the Youth Innovation Promotion Association (Grant No. 2022YSBR-048).

\bibliography{reference}

%%%%%%%%%% Merge with supplemental materials %%%%%%%%%%
%\pagebreak
%\newpage
\clearpage
\onecolumngrid
\begin{center}
\textbf{\large Supplemental Material: A phenomenological approach to Cupartes superconductors}
\end{center}

%%%%%%%%%% Merge with supplemental materials %%%%%%%%%%
%%%%%%%%%% Prefix a "S" to all equations, figures, tables and reset the counter %%%%%%%%%%
\setcounter{equation}{0}
\setcounter{figure}{0}
\setcounter{table}{0}
\setcounter{page}{1}
\makeatletter
\renewcommand{\theequation}{S\arabic{equation}}
\renewcommand{\thefigure}{S\arabic{figure}}
\renewcommand{\bibnumfmt}[1]{[S#1]}
\renewcommand{\citenumfont}[1]{S#1}
%%%%%%%%%% Prefix a "S" to all equations, figures, tables and reset the counter %%%%%%%%%%

\twocolumngrid

\subsection{Interaction channels}
\begin{figure}
	\begin{center}
		\fig{3.4in}{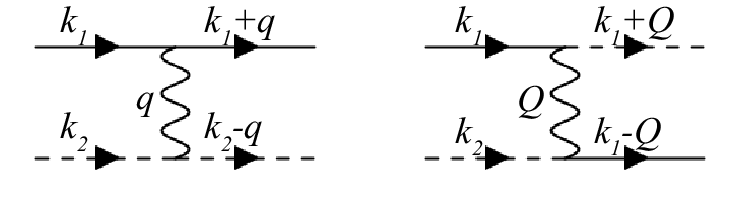}
		\caption{The Feynman diagram for interactions. $q$ stands for the small momentum transfer while the $Q$ stands for the large momentum transfer.
			\label{inter}}
	\end{center}
	%	\vskip-0.5cm
\end{figure}

In this subsection, we focus on the interaction forms.
Owing to the momentum conservation and without considering the Umklapp interaction, there are two interaction channels with small momentum transfer q and large momentum transfer Q, as shown in  Fig.\ref{inter}. More precisely, the interactions can be written as
\begin{eqnarray}
 u_q c_{k_1+q,\alpha}^\dagger c_{k_2-q,\beta}^\dagger c_{k_2,\beta}c_{k_1,\alpha}+u_Q c_{k_1+Q,\alpha}^\dagger c_{k_2-Q,\beta}^\dagger c_{k_2,\beta}c_{k_1,\alpha}
\end{eqnarray}
where the $\alpha$, $\beta$ are spin indices. $k_1$, $k_2$ can be any momentum inside the Brillouin Zone.
If we focus on $k$ and $-k$, we will have 
\begin{eqnarray}
    &&u_1c_{k,\alpha}^\dagger c_{k,\beta}^\dagger c_{k,\beta}c_{k,\alpha}  \\
    &+&u_2 c_{k,\alpha}^\dagger c_{-k,\beta}^\dagger c_{-k,\beta}c_{k,\alpha} \\
    &+&u_3 c_{-k,\alpha}^\dagger c_{k,\beta}^\dagger c_{-k,\beta}c_{k,\alpha}
\end{eqnarray}
$u_{1/2}$ comes from the $u_q$ with $q=0$. $u_3$ comes from the large Q with $Q=-2k$. Grouping these terms into a compact form, we will arrive at the general interaction in Eq.\ref{hk}. The general interaction is an extension of the two-orbital Hubbard model. One caveat here is there may be one double counting. The part of the above interactions belongs to gap function interaction, which should be removed in principle.

The remaining part of $u_{q/Q}$ interactions are the residue interactions in $H_{res}$.

\subsection{Excitation and Operatros}
After constructing the ground state wavefunction,
\begin{eqnarray}
	|GS \rangle={\displaystyle \prod_k}'(u_k+d_kb_k^{\dagger}b_{-k}^{\dagger}+\frac{v_k}{\sqrt{2}}(b_k^{\dagger}+b_{-k}^{\dagger}))|0\rangle
\end{eqnarray}
the central task is to determine its excitations. Applying the electron operators to the $|GS \rangle$, we have
\begin{eqnarray}
    c_{k\uparrow}|GS\rangle&=&(\frac{v_{k}}{\sqrt{2}}c_{-k\downarrow}^{\dagger}+d_{k}c_{-k\downarrow}^{\dagger} b_{-k}^{\dagger}) |GS'\rangle \label{excitation1_1}\\
    c_{k\uparrow}^\dagger|GS\rangle&=&(u_{k}c_{k\uparrow}^\dagger+\frac{v_{k}}{\sqrt{2}}c_{k\uparrow}^\dagger  b_{-k}^{\dagger})|GS'\rangle \\
    c_{k\downarrow}|GS\rangle&=&(-\frac{v_{k}}{\sqrt{2}}c_{-k\uparrow}^{\dagger}-d_{k}c_{-k\uparrow}^{\dagger} b_{k}^{\dagger}) |GS'\rangle \label{excitation1_2}\\
    c_{k\downarrow}^\dagger|GS\rangle&=&(u_{k}c_{k\downarrow}^\dagger+\frac{v_{k}}{\sqrt{2}}c_{k\downarrow}^\dagger  b_{k}^{\dagger})|GS'\rangle
    \label{excitation4}
\end{eqnarray}
Hence, the electron operators $c$ is related to the mixing between single-particle operator and three-particle operator like $c_{k\uparrow}\sim \frac{v_{k}}{\sqrt{2}}c_{-k\downarrow}^{\dagger}+d_{k}c_{-k\downarrow}^{\dagger} b_{-k}^{\dagger}$. This is the essential property of correlated pairing wavefunction. On the other hand, the electron operators in the BCS wavefunction are related to single-particle operators as
\begin{eqnarray}
     c_{k\uparrow}|GS\rangle_{BCS}=v_{k}c_{-k\downarrow}^{\dagger} |GS'\rangle_{BCS} \\
     c_{k\uparrow}^\dagger|GS\rangle_{BCS}=u_{k}c_{k\uparrow}^{\dagger} |GS'\rangle_{BCS}
\end{eqnarray}

Generally speaking, the single-particle operator and three-particle operator applied states are no-longer eigenstates in a correlated system, which are difficult to determine in common. In the simplified HK model with pairing, they coupled through the pairing $\Delta_k$ into a 2*2 matrix in Eq. \ref{hodd}. Then, the odd electron excitation states are diagonalizable. For example,
\begin{eqnarray}
    |O_{k+}^1\rangle&=& x_k c_{k\uparrow}^\dagger|0\rangle+y_k b_{-k}^\dagger c_{k\uparrow}^\dagger |0\rangle \\
    |O_{k-}^1\rangle&=& -y_k c_{k\uparrow}^\dagger|0\rangle+x_k b_{-k}^\dagger c_{k\uparrow}^\dagger |0\rangle
\end{eqnarray}
where the $x_k$, $y_k$ are Bogoliubov transformation factors from Eq. \ref{hodd}. These features are listed in Table.\ref{table}.

The electron operators can be rewritten in terms of Hubbard operators.
\begin{eqnarray}
c_{k\uparrow}^\dagger&=&
   a_{k+}^1 |O_{k+}^1\rangle \langle GS|+a_{k-}^1 |O_{k-}^1\rangle \langle GS| \\
   &&+ a_{k+}^4 |GS\rangle \langle O_{k+}^4|+a_{k-}^4|G.S.\rangle \langle O_{k-}^4| \\
c_{k\downarrow}^\dagger&=&
   a_{k+}^2 |O_{k+}^2\rangle \langle GS|+a_{k-}^2 |O_{k-}^2\rangle \langle GS|  \\
   &&+ a_{k+}^3 |GS\rangle \langle O_{k+}^3|+a_{k-}^3|GS\rangle \langle O_{k-}^3| 
\end{eqnarray}
The coefficients $a_{k\pm}^i$ can be easily determined both analytically and numerically. Therefore, the $c_{k\alpha}^\dagger$ are mapped to four low-energy excitation states as we claimed in the main text. For the comparison, the electron operators in BCS are
\begin{eqnarray}
    c_{k\uparrow}^\dagger&=&u_k |\phi_{k\uparrow}\rangle \langle GS|_{BCS}+v_k |GS\rangle_{BCS} \langle \phi_{-k\uparrow}| \\
    c_{-k\downarrow}^\dagger&=& u_k |\phi_{-k\downarrow}\rangle \langle GS|_{BCS}-v_k |GS\rangle_{BCS} \langle \phi_{k\uparrow}| 
\end{eqnarray}
where $|\phi_{k\uparrow}\rangle / |\phi_{-k\downarrow}\rangle $ are the normalized one-quasi-particle states \cite{schrieffer}.
After obtaining the excitations, the retarded Green's functions for operator $\hat{A}$, $\hat{B}$ can be easily found using Lehmann representation.
\begin{equation}
    \langle\langle A|B\rangle\rangle(\omega)=\sum_n \frac{\langle GS| \hat{A} |n \rangle \langle n| \hat{B} |GS \rangle}{\omega+i\eta-E_{n0}}+\frac{\langle GS| \hat{B} |n \rangle \langle n| \hat{A} |GS \rangle}{\omega+i\eta+E_{n0}} 
    \label{Lehmann}
\end{equation}
where $|n \rangle$ is the eigenstate for $H$ with eigenvalue $E_n$ and $E_{n0}$ is the energy difference between $|n \rangle$ and $|GS \rangle$.

There is one important caveat. Although we use the HK interaction to determine our excitaions, the arguments here only depende on the wavefunction $|GS\rangle$. One can hardly imagine the mixing between single-particle operator and three-particle operator in the above excitaion states are still single-particle. The current operators must mapped to more states than the BCS cases. The Green's functions must contain at least four poles.

\subsection{Exact diagonalization}
To demonstrate the particle-hole asymmetry, we apply the exact diagonalization (ED) method using the QuSpin package \cite{quspin1,quspin2}. Due to the limit of ED, we take a paired 1D fermionic chain as an example. The Hamiltonian is defined as
\begin{equation}
    H_{1D}=\sum_{\langle ij\rangle}t c_{i\sigma}^{\dagger} c_{j\sigma}-\Delta_0 (c_{i\uparrow}^{\dagger} c_{j\downarrow}-c_{i\downarrow}^{\dagger} c_{j\uparrow})+h.c. +U \hat{n}_{i\uparrow}\hat{n}_{i\downarrow}
\end{equation}
The chemical potential is tuned to $1.0$ for $k_F=2\pi/3$, which is accessible in the 12-sites 1D nearest-neighbor fermionic chain. The pairing function becomes the extended $s$-wave pairing function $\Delta_0\cos k_x$ and we choose $\Delta_0=0.2$. The $A(k,\omega)$ spectrum can be easily obtained from ED and has been implemented in the Quspin. When $U=0$, the $A(k_F,\omega)$ is symmetric as shown in Fig.\ref{fig2}(c) as expected from BCS. On the contrary, when we add a small Hubbard interaction $U$, the $A(k_F,\omega)$ becomes asymmetric although the energies remain symmetric, as shown in Fig.\ref{fig2}(d) using $U=0.2$.

\subsection{Superfluid density}
The superfluid density $n_s$ can be obtained from the linear response theory according to Eq.\ref{ns}. The part of the paramagnetic contribution is obtained by the current-current correlation function
\begin{eqnarray}
    \Pi_{\mu \nu}(q,\omega)&=&-\frac{i}{V\hbar}\int_{0}^{\infty}dt e^{i\omega t} \langle [J_\mu(q,t),J_\nu(-q,0)]\rangle \nonumber
\end{eqnarray}
which can be calculated by analytic continuation $i\omega_n\rightarrow\omega+i\eta$ of
\begin{equation}
    \Pi_{\mu \nu}(q,i\omega)=\frac{1}{V\beta}\sum_{k,i\nu_n} \frac{\partial\epsilon_{k+\frac{q}{2}}}{\partial_{k_\mu}} \frac{\partial\epsilon_{k+\frac{q}{2}}}{\partial_{k_\nu}} \mathrm{Tr}\left[\bar{\mathcal{G}}(k,i\nu_n)\bar{\mathcal{G}}(k+q,i\nu_n+i\omega_n)\right]
\end{equation}
where $\epsilon_k=-2t(\cos k_x +\cos k_y)$ is the dispersion of the normal state and $\bar{\mathcal{G}}(k,i\nu_n)$ is the Green's function under Nambu's basis in which the normal and anomalous part is obtained using Lehmann representation. Actually, the correlation function can also be calculated directly using Lehmann representation Eq.\ref{Lehmann} just by replacing $\hat{A}$ and $\hat{B}$ with current operator Eq.\ref{current}. The results obtained by these two methods are consistent as expected. After some direct but complicated calculation, we get the results showed in Fig.\ref{fig3}. \par
Since the current operator describe the particle-hole excitation, the matrix element in Lehmann representation means the overlap between the particle-hole excitation from $\ket{GS}$ and the excited eigenstates in single-electron subspace, and if explicitly written, it will contain the term like
\begin{equation}
    \frac{\mel{GS}{c_{k'\uparrow}^{\dagger}c_{k'+q\uparrow}}{O^i_\pm O^j_\pm}_{k+q,k} 
    \mel{O^i_\pm O^j_\pm}{c_{k+q\uparrow}^{\dagger}c_{k\uparrow}}{GS}}{\omega+i\eta-E_{n0}}.
\end{equation}
In our calculation, we find that the contribution of particle-hole excitation to the same excited states like $|O_{k-}^1O_{k+}^1\rangle$ vanishes, just as the same as that in the BCS theory. However, the contribution of different excited states like $|O_{k-}^1O_{k+}^4\rangle$ is not zero, which results in the reduction of superfluid density. Note that these non-zero matrix elements are due to the muti-pole structure in Green's function caused by the interaction.

\subsection{Impurity scattering}
To investigate the effect of impurity scattering, we consider the following impurity Hamiltonian,
\begin{equation}
    H_{imp} = \sum_{il} V_0 \delta(\br_i-\mathbf{R}_l)c_i^\dg c_i = \frac{1}{V}\sum_{\bk \bq l} V_0 e^{i \bq\cdot\mathbf{R}_l} c_{k+q}^\dg c_k 
\end{equation}
in which $\mathbf{R}_l$ is the position of impurities, $V_0$ is the strength of impurity potential and we only consider the $\delta$ scattering potential for convenience. The effects of impurities can be calculated by perturbation expansion of $H_{imp}$. The Dyson equation is
\begin{equation}
    G(\bk,\omega) = G^{(0)}(\bk,\omega) + G^{(0)}(\bk,\omega)\Sigma(\omega)G(\bk,\omega)
\end{equation}
where the self-energy needs to be determined through self-consistent calculation. Note that $G^{(0)}(\bk,\omega)$ is just Green's function of non-interacting electrons for $U_k=0$. For $U_k\neq0$, we expect that such perturbation expansion is still appropriate if $V_0$ is not too large. Then, we replace the  non-interacting Green’s function obtained through Lehmann's representation. \par
For random impurities, we need to average over $\mathbf{R}_l$. When the density of impurities $n_{imp}=\frac{N_{imp}}{V}$ is small, we can ignore the scattering process on different impurities which is called the full Born approximation. The Feynman diagram is showed in Fig.\ref{fig4}(a) and it's straightforward to write down the n-th order self-energy 
\begin{equation}
    \Sigma^{(n)}(\omega)=n_{imp}V_0 \sigma_3 ( V_0\sigma_3 G(\omega))^{n-1}
\end{equation}
in which $\sigma_3$ is the Pauli matrix in particle-hole space and $G(\omega)=\frac{1}{V}\sum_\bk G(\bk,\omega)$. Then the total self-energy is
\begin{equation}
    \Sigma(\omega)=\frac{n_{imp}V_0\sigma_3}{1-V_0\sigma_3 G(\omega)}.
\end{equation}
After the self consistent calculation on $G(\bk,\omega)$ and $\Sigma(\omega)$, we get DOS from the impurity scattering modified Green's function.

\end{document}